\documentclass[twocolumn,notitlepage,superscriptaddress, nofootinbib,nobibnotes, aps,prd,10pt]{revtex4-1}
\usepackage{amsmath,amssymb, fp}  
\usepackage{xcolor}  
\usepackage{soul} 
\usepackage{booktabs}
\definecolor{linkcolor}{RGB}{150,50,60}  
\usepackage[	colorlinks=true,    	
			pdfstartview=FitV,
			linkcolor= linkcolor,
			citecolor= linkcolor,
			urlcolor= linkcolor,
			linktoc=page,
			hyperindex=true,
			hyperfigures=true,  
			breaklinks=true]
			{hyperref}
\usepackage{lastpage}  
\usepackage{graphicx}
\usepackage[caption=false]{subfig}
\usepackage{placeins}

\newcommand{\up}{\mathrm{up}}
\newcommand{\out}{\mathrm{out}}
\newcommand{\dd}{\mathrm{d}}
\newcommand{\kk}{\kappa}
\newcommand{\EF}{\mathcal E}
\newcommand{\ee}{\mathbf e}

\begin{document}

\title{Energy extraction via magnetic reconnection in slowly spinning Kerr black holes:\\ \smallskip The role of the magnetosphere }
\author{Jibril Ben Achour}
\affiliation{Univ de Lyon, ENS de Lyon, Laboratoire de Physique, CNRS UMR 5672, Lyon, France}
\affiliation{Institut de Math\'ematiques de Bourgogne, Univ. Bourgognes Europe, Dijon, France}
\author{Ileyk El Mellah}
\affiliation{Departament de Física, EEBE, Universitat Politècnica de Catalunya, c/Eduard Maristany 16, 08019 Barcelona, Spain}

\begin{abstract}
Magnetic reconnection in black hole magnetospheres has become a promising mechanism for powering bright flares from accreting black holes. GR-MHD and GR-PIC numerical simulations have established that reconnection is often accompanied by the fragmentation of current sheets triggering the formation of plasmoids, independently of the black hole spin. Within the ergoregion, some of these plasmoids may acquire negative Killing energy, opening a channel for the extraction of the black hole rotational energy. However, current analytical models predict that the production of negative-energy plasmoids generally requires black hole spins above $a\sim 0.8$ (for realistic magnetization). Here, we generalize existing analytical models by incorporating, for the first time, the effects of the Blandford-Znajek (BZ) magnetosphere on the plasma dynamics. We analytically show that the drift imposed by the BZ magnetosphere on the upstream plasma allows the formation of negative-energy plasmoids for arbitrarily small (but non-vanishing) spins and magnetization. Our results highlight the crucial role of the magnetospheric structure in determining the energetics of reconnection-generated plasmoids and provide a more realistic framework for bridging the gap between analytical models and numerical simulations.

\end{abstract}
\maketitle

Black hole are well known to exhibit a rich activity which can be split into a quiescent emission supplemented by bright transient episodes \cite{Falcke:2013ola}. For Sagittarius $A^{\ast}$, the later correspond to flares in the near-infrared and X-rays bands which generically last for few hours. The ESO-VLTI/GRAVITY collaboration reported the observation of three such events, successfully monitoring the kinematics of the hotspot trajectory \cite{GRAVITY:2020hwn, GRAVITY:2020lpa, GRAVITY:2023avo}. Combined astronometric and polarimetric measurements allow us to constrain the radius and period of the hot spot orbit, the speed of its centroid, the deviation from Keplerian motion as well as the geometry of the magnetic field in this region. Yet, the mechanism sourcing these bright flares is still poorly understood. In the recent years, the fast reconnection of magnetic fields lines in the vicinity of the collisionless black hole magnetosphere have been recognized as one key potential mechanism to explain these flares. Since then, important efforts have been devoted, both numerically and analytically, to better characterize to which extent the magnetic reconnection provides a astrophysically efficient process to i) reproduce the flaring activity and ii) to extract energy from a spinning magnetized black hole \cite{Matsumoto:2020wul, Antonopoulou:2024qco, Yfantis:2024eab, Vincent:2023sbw, Aimar:2023kzj, Parfrey2019, Ripperda:2020bpz, Ripperda:2021zpn, Bransgrove:2021heo, ElMellah:2021tjo, Nathanail:2021jbn, ElMellah:2023sun, Jiang:2023ygs, Vos:2023ska, Meringolo2025, Xia:2026nkt, ComissoAsenjo2021, Chen:2024ggq, Zeng:2026ufr, YuChih:2026ios, Shen:2024sdr, Wan2026, Camilloni:2024tny, BenAchour:2026gua}.

Reconnection of magnetic fields lines can be triggered in various physical configurations, including the accumulation of magnetic flux on the black hole horizon in magnetically arrested disc, or the reversal of the magnetic field across the equator in the split-monopole black hole magnetosphere. A common features of all these scenarios is the presence of unstable current sheets.
Within these layers, the ideal plasma conditions can break down, allowing magnetic field lines to reconnect and electromagnetic energy to be transferred to the plasma. This mechanism is observed in general relativistic magnetohydrodynamics (GR-MHD) and particle-in-cell (GR-PIC) simulations, where the tearing instability induces a fragmentation of the current sheets and the formation of magnetized plasmas structures, dubbed plasmoids, which are either accreted by the hole or ejected from its vicinity \cite{Ripperda:2020bpz, Ripperda:2021zpn, Bransgrove:2021heo, ElMellah:2021tjo, Nathanail:2021jbn, ElMellah:2023sun, Jiang:2023ygs, Vos:2023ska, Meringolo2025, Xia:2026nkt}. If one of the plasmoids carries negative Killing energy and is absorbed by the hole, it can contribute to a Penrose-like extraction of rotational energy. Either the second positive-energy plasmoid escapes to infinity, or it can orbit around the hole and transfer part of its energy to the surrounding plasma or through radiation before getting accreted too. In both case, the process can extract energy to the hole. Therefore, understanding the condition to generate negative-energy plasmoids through magnetic reconnection in the Kerr black hole has attracted much attention recently \cite{ComissoAsenjo2021, Chen:2024ggq, Zeng:2026ufr, YuChih:2026ios, Shen:2024sdr, Wan2026}. 


The seminal model for the generation of negative-energy plasmoids via magnetic reconnection on the Kerr black hole was proposed by Comisso and Asenjo, relying on the minimal simplifying assumption that the upstream plasma before reconnection follows a Keplerian trajectory, i.e. a circular equatorial geodesic. Under this assumption, the generation of negative-energy plasmoid via magnetic reconnection is allowed only above some critical spin threshold given by $a_{\ast}\sim 0.9$ \cite{ComissoAsenjo2021}. More recently, generalizations of this model have been constructed, based on circular or plunging geodesic motion for the upstream plasma. In general, these models require rapidly rotating black holes \cite{Chen:2024ggq, Zeng:2026ufr, YuChih:2026ios}, i.e with $a > 0.8$, but the production of negative-energy plasmoids is also possible at lower spins although it requires increasingly large values of the magnetization $\sigma_0$ \cite{Shen:2024sdr}. 


In parallel, GR-MHD and GR-PIC simulations have provided evidence for the production of negative-energy particles in the ergosphere of black hole magnetospheres. While most of them focus on high spin values \cite{Parfrey2019, ElMellah:2021tjo, Xia:2026nkt}, this phenomenon has also been recently reproduced for spin below the spin threshold found analytically. In particular, the recent simulations of a split-monopole magnetosphere by Meringolo et al. has revealed populations of positrons with negative Killing energy inside the ergoregion already at $a=0.7$, together with plasmoid formation in the equatorial current sheet \cite{Meringolo2025}. Although the negative energy of individual particles does not establish that a whole plasmoid carries negative energy, these results motivate revisiting the apparent high-spin restriction of existing analytical models. In particular, they raise the question of how the local production of negative-energy plasmoids depends on the upstream plasma motion imposed by the black hole magnetosphere.

In this letter, we demonstrate that, provided magnetic reconnection takes place, the production of a negative-energy plasmoid from a positive-energy upstream plasma is kinematically allowed at a \textit{arbitrarily small nonzero spin} and \textit{without requiring irrealistically large magnetizations}. This result crucially relies on taking into account the Blandford--Znajek (BZ) split-monopole magnetosphere \cite{BlandfordZnajek1977}, which induces a drift of the upstream plasma in the plane orthogonal to the magnetic field. This changes the local energy criterion compared with prescriptions based on geodesic upstream motion. 
This letter is organized as follows. We first introduce the slowly rotating BZ background and recast the key quantities in the appropriate streched coordinate $\kappa$. Then, we fix the upstream plasma frame and derive the drift, Eq~(\ref{eq:drift}), of the upstream plasma imposed by the magnetosphere. Using these quantities, we derive the extent of the equatorial region, Eq~(\ref{eq:inequality}), which allows the co-existence of a positive energy in-state and a one negative-energy out-state assuming an instantaneous magnetic reconnection. The new result generalizes previous works by taking into account the ejection direction w.r.t i) the BZ magnetic field and ii) the upstream drift direction. Finally, we confront our result to the previous works on energy extraction via magnetic reconnection and in particular the need for an escape condition usually imposed to the second plasmoid.

\section{Blandford-Znajek background and the reconnection layer}
\label{sec:background}

In the following we consider a magnetized Kerr black hole of mass $M$ and \emph{dimensionless} spin
$a=J/M^2>0$. We shall work in the slow spin regime such that $a\ll1$ and $G=c=1$.  In Boyer--Lindquist coordinates, the exact Kerr geometry reads
\begin{equation}
\begin{aligned}
\dd s^2={}&-\alpha^2\dd t^2+\frac{\Sigma}{\Delta}\dd r^2
 +\Sigma\dd\theta^2+\varpi^2(\dd\phi-\omega\dd t)^2
\end{aligned}
\label{eq:kerr}
\end{equation}
where the functions $(\Sigma, \Delta, A)$ entering in the metric are given by $\Sigma=r^2+a^2M^2\cos^2\theta$, $\Delta =\Delta=r^2-2Mr+a^2M^2$ and $A=(r^2+a^2M^2)^2-a^2M^2\Delta\sin^2\theta$ while the other functions $(\alpha, \varpi, \omega)$ read
\begin{align}
\alpha^2=\frac{\Delta\Sigma}{A},\quad \varpi^2=\frac{A\sin^2\theta}{\Sigma}, \qquad \omega=\frac{2aM^2r}{A}.
\end{align}
Let us also introduce the combination $\beta^{\hat\phi} = \omega\varpi / \alpha$ which will appear useful later on.
The position of the outer Kerr horizon $r_{+}$ and outrer ergosphere $r_{E}$ are given by
\begin{align}
r_{+} & = M(1+\sqrt{1-a^2}) \\
 r_{E} &= M(1+\sqrt{1-a^2\cos^2\theta})
\end{align}
In order to model the effect of the Kerr magnetosphere, we consider the surrounding plasma as embedded in the stationary, axisymmetric, force-free magnetosphere $(F, J)$. Such force-free Maxwell solution can be parametrized, on either side of the equatorial current sheet, by a Faraday tensor $F$ of the form
\begin{align}
F={}&\dd\Psi\wedge(\dd\phi-\Omega_F(\Psi)\dd t)-\frac{\Sigma\mathcal I(\Psi)}{\Delta\sin\theta}\dd r\wedge\dd\theta,
\label{eq:bzpotential}
\end{align}
where $\Psi(r,\theta)=\mathcal A_\phi$ is the magnetic flux function,
$\Omega_F(\Psi)$ is the angular velocity of the field lines, and
$\mathcal I(\Psi)=\sqrt{-g}F^{\theta r}$ is the poloidal current function in our conventions. To fix a realistic model, we choose the slowly rotating Blandford-Znajek split monopole \cite{BlandfordZnajek1977}. At second order in the spin expansion, the BZ functions $(\Psi, \Omega, \mathcal  I)$ are given by
\begin{align}
\Psi&=\Psi_0(1-|\cos\theta|)+O(a^2\Psi_0),\nonumber\\
\Omega_F&=\frac{a}{8M}+O(a^3/M),\nonumber\\
\mathcal I&=\frac{a\Psi_0}{8M}\operatorname{sgn}(\cos\theta)
\sin^2\theta+O(a^3\Psi_0/M).
\label{eq:bzfunctions}
\end{align}
The current function is defined separately in each hemisphere, with opposite signs across the current sheet. The corresponding electromagnetic four-current $J^\mu \partial_{\mu}$ follows from Maxwell's equations $\nabla_{\mu}F^{\mu\nu} = 4\pi J^{\nu}$ but we will not need its expression in what follows. However, being a force-free solution, it satisfies $F_{\mu\nu} J^{\nu} =0$. This provides us with a concrete model to study how the Kerr magnetosphere affects the magnetic reconnection process. 

Let us introduce a ZAMO observer with four-velocity $n=\alpha^{-1}(\partial_t+\omega\partial_\phi)$ and orthonormal spatial triad
\begin{equation}
\begin{aligned}
 e_{\hat r}&=\sqrt{\Delta/\Sigma}\,\partial_r,\quad e_{\hat\theta}&=\Sigma^{-1/2}\partial_\theta,\quad
 e_{\hat\phi}=\varpi^{-1}\partial_\phi.
\end{aligned}
 \label{eq:zamo}
\end{equation}
In this frame $(n, \ee_{\hat r}, \ee_{\hat\theta}, \ee_{\hat\phi})$, the leading equatorial BZ fields in the stretched layer defined below decompose as
\begin{equation}
\begin{aligned}
\mathbf B&=B^{\hat r}(\ee_{\hat r}+v_F\ee_{\hat\phi})+O(a^2 B^{\hat r}),\\ \mathbf E&=-v_F B^{\hat r}\ee_{\hat\theta}+O(a^2 B^{\hat r})
\end{aligned}
\label{eq:fields}
\end{equation}
where $v_F$ is given by 
\begin{align}
v_F&\equiv\frac{\varpi(\Omega_F-\omega)}{\alpha} \\
&= \frac{a r \sin\theta}{8M\sqrt{1-2M/r}} \left( 1 -\frac{16 M^3}{r^3}\right) + O(a^3)
\end{align}
The second equality is a fixed-radius expansion valid for $r>2M$ only. Here $v_F$ is the velocity of a magnetic surface relative to the ZAMO,
which should not be confused with the azimuthal plasma velocity.  At this stage, it is interesting to point that $B^{\hat\phi}=-E^{\hat\theta}+O(a^2 B^{\hat r})$. This relation  follows from the future-horizon (Znajek) regularity
in the region studied below. See Appendix-\ref{app:bz} for more details.

Now, it is useful to introduce adapted coordinates to our problem. Following \cite{ComissoAsenjo2021}, we assume that the reconnection point lies in the equatorial plane, i.e. $\theta=\pi/2$, and within the ergoregion but outside the horizon, i.e. $r_{+} < r < r_E$.
At the equator and in the slowly rotating regime, the position of the outer horizon and the ergoregion become respectively $r_+=2M-a^2M/2+O(a^4M)$ and $r_E=2M$, allowing for a tiny region $\Delta r = a^2M/2$.  To resolve this shrinking region, it will be convenient to introduce the stretched coordinate $\kappa$ defined as
\begin{equation}
\begin{gathered}
r_X=2M-\kk a^2M\quad \text{with} \qquad 0<\kk<\tfrac12,
\end{gathered}
 \label{eq:layer}
\end{equation}
We use $\kappa$ as the sole stretched radial coordinate throughout. 
At fixed $\kk$ in the slow-spin limit, the metric functions $(\omega, \varphi)$ become constant at that order, i.e.
\begin{equation}
\begin{aligned}
\omega=\frac{a}{4M}+O(a^3/M),\quad \varpi&=2M+O(a^2M)
\end{aligned}
\label{eq:layergeom}
\end{equation}
while the functions $(\alpha, \beta^{\hat\phi})$ are given by
\begin{equation}
\begin{aligned}
\alpha & =\frac{a\sqrt{1-2\kk}}{2}+O(a^3),\\ \beta^{\hat\phi}&\equiv\frac{\omega\varpi}{\alpha}
=\frac1{\sqrt{1-2\kk}}+O(a^2).
\end{aligned}
\label{eq:layerlapse}
\end{equation}
Instead, inserting $r_X$ into the exact Kerr functions before expanding gives $\Delta(r_X)=M^2[a^2(1-2\kk)+a^4\kk^2]$. Using $\Omega_F-\omega=-a/(8M)+O(a^3/M)$ and the lapse above, one obtains 
\begin{align}
v_F&\equiv\frac{\varpi(\Omega_F-\omega)}{\alpha}
=-\frac{1}{2\sqrt{1-2\kk}}+O(a^2).
\end{align}
Independently, the current function in Eq.~\eqref{eq:bzfunctions} fixes the toroidal field through the exact identity $B^{\hat\phi}=-\mathcal I/(\sqrt\Delta\sin\theta)$. On approaching the equator from the northern hemisphere, at fixed $\kk$, we obtain
\begin{align}
 B^{\hat r}&=\frac{\Psi_0}{4M^2}+O(a^2\Psi_0/M^2),\nonumber\\
 B^{\hat\phi}&=-\frac{\Psi_0}{8M^2\sqrt{1-2\kk}}+O(a^2\Psi_0/M^2),\nonumber\\
 E^{\hat\theta}&=\frac{\Psi_0}{8M^2\sqrt{1-2\kk}}+O(a^2\Psi_0/M^2).
\label{eq:bzlayerfields}
\end{align}
All three fields reverse sign on the southern side. Thus $B^{\hat\phi}/B^{\hat r}=v_F+O(a^2)$, confirming Eq.~\eqref{eq:fields} directly from the BZ current, without using the fixed-radius expansion inside the ergoregion. These estimates assume regular subleading BZ flux corrections and fixed $0<\kk<1/2$. This provides us with a complete description of the region of interest in our magnetized Kerr black hole. We can now discuss the initial and final configuration of the plasma, before and after the reconnection process.

\section{Local frames and the upstream state}
\label{sec:frames}

To be concrete, we consider a plasma orbiting in the dynamical Kerr magnetosphere.  The energy-momentum tensor of the system is given by
\begin{align}
\label{en}
T_{\mu\nu} = \left( \rho + P\right)  U_{\mu} U_{\nu} + P g_{\mu\nu} + F_{\mu\alpha} F^{\alpha}{}_{\nu} - \frac{1}{2} g_{\mu\nu} F_{\alpha\beta} F^{\alpha\beta} \nonumber
\end{align}
and we assume that we can neglect the backreaction. We assume that the magnetosphere is described by the stationary and axi-symmetric BZ solution $F^{BZ}$ supplemented with rapidly varying perturbations $F^{\text{MR}}_{\mu\nu}$ responsible for the magnetic reconnection such that the full Faraday tensor reads
\begin{align}
F_{\mu\nu} = F^{\text{BZ}}_{\mu\nu} + F^{\text{MR}}_{\mu\nu}
\end{align}
Following \cite{ComissoAsenjo2021}, we do not assume any specific model for $F^{\text{MR}}_{\mu\nu}$ and only focus on the state of the plasma before and after reconnection where $F^{\text{MR}}_{\mu\nu} =0$. Let us now introduce the first key effect of the magnetosphere.

Before reconnection, the motion of the upstream plasma is constrained to follow a minimal ideal MHD flow such that $F^{\text{BZ}}_{\mu\nu} U^{\nu} =0$. Within the ZAMO frame $(n, \ee_{\hat r}, \ee_{\hat\theta}, \ee_{\hat\phi})$, this condition forces the plasma to drift orthogonally to the magnetic field $\mathbf B$, fixing the motion of the bulk plasma in this plane. This is the key novelty triggered by turning on the Kerr magnetosphere. Denoting the upstream four velocity in the ZAMO frame as
\begin{equation}
\begin{aligned}
U_{\up}^{\hat a}=\gamma_D(1,\mathbf v_D).
\end{aligned}
\label{eq:drift}
\end{equation}  
where the subscript $D$ denotes the drift. A direct computation of $(\gamma_D, \mathbf v_D)$ gives us
\begin{equation}
\begin{aligned}
\gamma_D&=\frac{\sqrt{5-8\kk}}{2\sqrt{1-2\kk}}, \\
\mathbf v_D&\equiv\frac{\mathbf E\times\mathbf B}{B^2}
=\frac{-\ee_{\hat r}-2\sqrt{1-2\kk}\ee_{\hat\phi}}{5-8\kk},
\end{aligned}
\label{eq:drift}
\end{equation}
which gives an inward radial drift for the plasma. Notice that contrary to previous models where the plunging trajectory of the upstream plasma is \textit{chosen} \cite{Chen:2024ggq, Zeng:2026ufr}, here it is fully \textit{imposed} by the BZ magnetosphere. Let us also stress that ideal MHD condition used here to fix the drift does not enforce a motion parallel to the magnetic field. We choose to set it to zero in what follows for simplicity.

Now, it is useful to build another frame adapted to the motion of the upstream plasma. To do so, let us define the two space-like vectors orthogonal to $U_{\up}$ which read
\begin{equation}
\begin{gathered}
\begin{aligned}
 \mathbf b&=\frac{-2\sqrt{1-2\kk}\ee_{\hat r}+\ee_{\hat\phi}}{\sqrt{5-8\kk}},\quad \mathbf d=\frac{-\ee_{\hat r}-2\sqrt{1-2\kk}\ee_{\hat\phi}}{\sqrt{5-8\kk}},
\end{aligned}
\end{gathered}
\label{eq:comovingbasis}
\end{equation}
and let us introduce a spatial basis w.r.t $U_{\up}$ given by
\begin{equation}
\begin{gathered}
\begin{aligned}
 B^{\hat a}&=(0,\mathbf b),\quad T^{\hat a}=(0,\ee_{\hat\theta}),\quad D^{\hat a}=(\gamma_Dv_D,\gamma_D\mathbf d).
\end{aligned}
\end{gathered}
\label{eq:comovingbasis}
\end{equation}
where $v_D$ is given by 
\begin{equation}
\begin{gathered}
\begin{aligned}
 v_D&=\frac1{\sqrt{5-8\kk}}
\end{aligned}
\end{gathered}
\label{eq:comovingbasis}
\end{equation}
Notice that the vector $T$ is in the same direction as the leading electric field of the magnetosphere.
The vectors $(U_{\text{up}}, B, T, D)$ form an orthonormal basis which will reveal useful to discuss the different configurations of motion. 
The key point is that a spatial direction in the plasma rest
frame is not generally spatial in the ZAMO frame. The ZAMO and the plasma rest frames provide the relevant structures to analyze the reconnection process and impose the required conditions of energy on the initial ad final states of the plasma before and after reconnection. 

\section{In/out states and negative-energy production}
\label{sec:criterion}
\textbf{In state.} Let us first impose that the initial upstream plasma carries a positive Killing energy. Immediately before reconnection, the upstream plasma
has four-velocity $U_{\up}$ given by \eqref{eq:drift} such that its specific Killing energy is
$\EF_{\up}=-k\cdot U_{\up}$, with $k=\partial_t$ the stationary Killing vector of the Kerr geometry. Since
$k=\alpha(n-\beta^{\hat\phi}e_{\hat\phi})$, a direct computation gives 
\begin{equation}
\EF_{\up}=\alpha\gamma_D
\bigl(1+\beta^{\hat\phi}v_D^{\hat\phi}\bigr)
=\frac{a(3-8\kk)}{4\sqrt{5-8\kk}}+O(a^3).
\label{eq:ein}
\end{equation}
Thus an admissible positive-energy initial state requires
\begin{equation}
\EF_{\up}>0\qquad\Longleftrightarrow\qquad 0<\kk<\frac38.
\label{eq:upcondition}
\end{equation}
This is the first condition on the region of interest. Notice that the specific value $3/8$ is directly imposed by the BZ solution which fixes the drift velocity $v_D$ entering in $U_{\text{up}}$. Let us now look for the conditions on the final two-plasmoid state.

\textbf{Out states.}  During the magnetic reconnection, the upstream plasma is split into two oppositely expelled plasmoids. In the instantaneous upstream rest frame, we denote by
$v_{\out}\in[0,1]$ the common speed of these two plasmoids.
To remain general, we consider that the ejection take place along a unit direction $S$ living in the 3d subspace spanned by the basis $(B, T, D)$  which can be parametrized as follows
\begin{equation}
\begin{gathered}
 S=\cos\xi\,B+\sin\xi\cos\chi\,T +\sin\xi\sin\chi\,D,
\end{gathered}
\label{eq:direction}
\end{equation}
where the range of the angles $(\xi, \chi)$ is given by $ 0\le\xi\le\tfrac\pi2$ and $ -\pi<\chi\le\pi$.  Since $S\rightarrow-S$ merely interchanges the two plasmoids, it is sufficient to restrict $\xi$ to $[0,\pi/2]$ while retaining the full azimuthal range for $\chi$. A direct check shows that $U_{\up}\cdot S=0$ and $S^2=1$ as needed.  The angle $\xi$ measures the inclination of the ejection direction with respect to the magnetic field, with $\xi=0$ corresponding to field-aligned ejection and $\xi=\pi/2$ to ejection orthogonal to the magnetic field. The azimuthal angle $\chi$ specifies the orientation of the transverse component of the ejection direction. For $\chi=\{0,\pm\pi \}$, the ejection is tilted in the polar direction and remains orthogonal to the drift, whereas for $\chi=\pm\pi/2$, the ejection remains within the equatorial plane and is tilted along or against the drift, respectively.

Locally, the four velocity $U_{\text{up}}$ of the upstream plasma before reconnection and the four velocity of the two plasmoids are related by a Lorentz transformation 
\begin{equation}
\begin{gathered}
 U_\pm=\gamma_{\out}(U_{\up}\pm v_{\out}S),
\end{gathered}
\label{eq:outstates}
\end{equation}
where $ \gamma_{\out}=(1-v_{\out}^2)^{-1/2}$. In this picture, the magnetic reconnection happens instantaneously which allow us to relate the initial and final state through (\ref{eq:outstates}). We can now evaluate the energy carried by the two plasmoids which is given by $\EF_\pm=-k\cdot U_\pm$. 
A direct computation combining \eqref{eq:ein}--\eqref{eq:direction} gives
\begin{equation}
 \EF_\pm
 =\frac{a\gamma_{\out}}{4\sqrt{5-8\kk}}
 \bigl[3-8\kk\pm v_{\out}\mathcal A\bigr]
 +O(a^3).
\label{eq:eout}
\end{equation}
where $\mathcal A: =\mathcal A(\kk,\xi,\chi)$ is given by 
\begin{equation}
\mathcal A(\kk,\xi,\chi)
 \equiv2\cos\xi-\sqrt{5-8\kk}\sin\xi\sin\chi.
\label{eq:angularA}
\end{equation}
such that it contains the information on the ejection angles $(\xi, \chi)$. Several remarks are in order at this stage. First the function $\mathcal  A(\kk,\xi,\chi)$ is invariant under $\chi \rightarrow \pi-\chi$ such that one can restrict the ranges of the angle to $\xi \in [0,\pi/2]$ and $\chi \in [-\pi/2, \pi/2]$ which covers all configurations. Second, notice that a \emph{polar} tilt ($\chi=0$) replaces $v_{\out}$ by
$v_{\out}\cos\xi$ in \eqref{eq:eout}. Finally, notice that a small tilt opposite to the
plunging drift enhances the negative-energy projection, whereas a small
tilt toward the drift reduces it. We can now derive the key result.

Recall that a positive-energy in state ($\EF_{\up}>0$) imposes the constraint \eqref{eq:upcondition}. The sign of $\mathcal A$ determines which branch carries the lowest Killing energy: $\EF_-<\EF_+$ for $\mathcal A>0$, and $\EF_+<\EF_-$ for $\mathcal A<0$. For $\mathcal A=0$, both branches carry the same positive energy. Therefore, a positive-energy in state and one negative-energy out state coexist precisely when, to leading order,
\begin{equation}
 \frac{3- v_{\out}\,\bigl|\mathcal A(\kk,\xi,\chi)\bigr|}{8} <  \kk< \frac{3}{8}
\label{eq:inequality}
\end{equation}
Within the domain \eqref{eq:upcondition}, one has $|\mathcal A|\leq\sqrt{9-8\kk}<3$. Since $v_{\out}<1$ for massive plasmoids, the lower bound in \eqref{eq:inequality} is strictly positive, while the upper bound satisfies $3/8<1/2$. Whenever non-empty, the resulting negative-energy production region therefore lies entirely within the ergoregion and outside the event horizon, to the order considered. This condition establishes the local production of a negative-energy plasmoid, independently of the subsequent evolution of the two plasmoids. This is the key result of this work.

\begin{figure*}[!t]
\centering
\IfFileExists{figure3_closed_CA_vs_BZ.pdf}{%
\includegraphics[width=\textwidth]{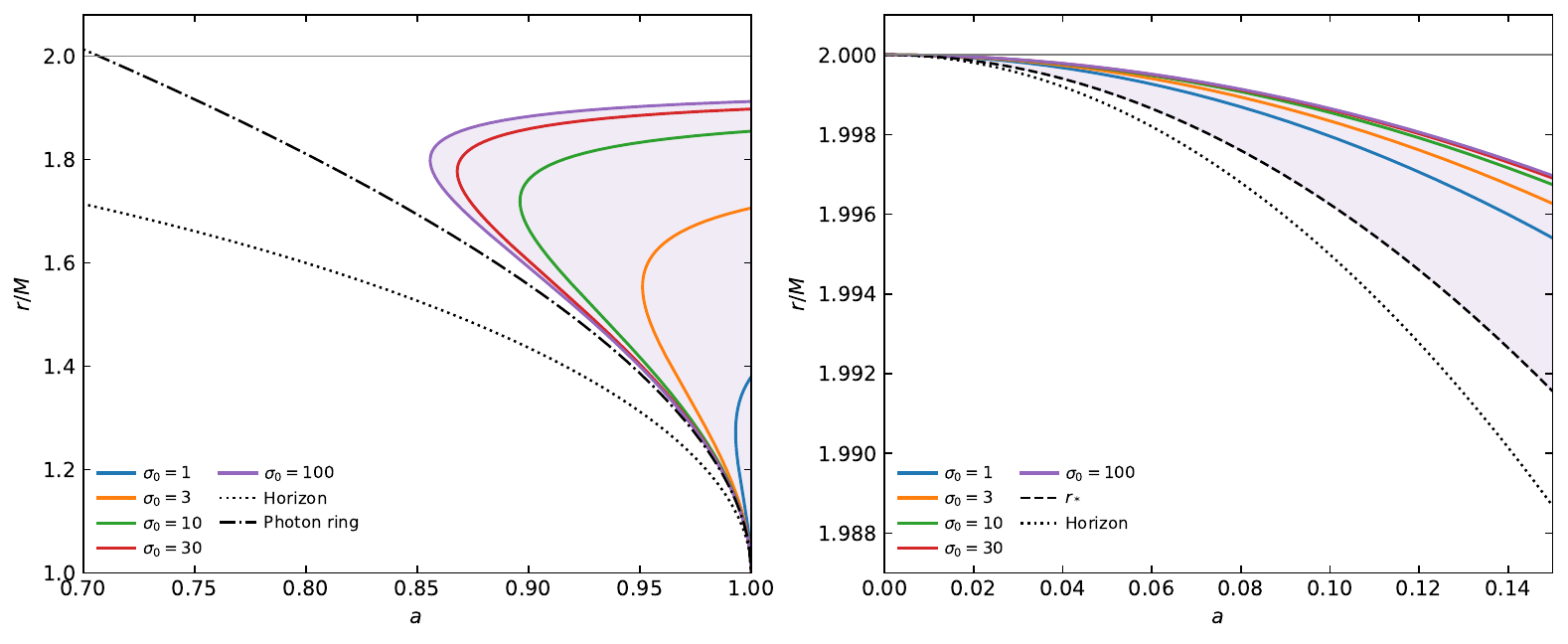}}{%
\fbox{\parbox[c][2.0cm][c]{0.1\textwidth}{\centering Original figure file not supplied: \\ \texttt{figure3\_closed\_CA\_vs\_BZ.pdf}}}}
\caption{\textbf{Local negative-energy production regions in the $(a,r/M)$ plane.}
Left: the original Comisso-Asenjo (CA) model, showing closed
negative-energy boundaries for several magnetizations and the
highlighted local production region. As manifest, the parameter space which allows BH rotational energy extraction (purple shaded region) only exists beyond a threshold spin whose value slowly depends on the plasma magnetization $\sigma_0$. Right: the slow-spin BZ model with
polar inclination $\xi=\pi/12$ and $\chi=0$. The upper curves are
$r_{\mathrm{neg}}(a,\sigma_0)$ and the dashed lower boundary is $r_*(a)$.
The dotted horizon and dash-dotted photon ring are shown for context. Notice that on the right panel, the photon ring cannot be represented because of the zoom used to represent that region.
The BZ curves are the leading-order $O(a^2)$ radial predictions and
are plotted only for $0\le a\le0.15$. In contrast to the CA model, the allowed region exists at arbitrary low spins.}
\label{fig:ca-bz-analytic}
\end{figure*}

A second important feature of this result is that the absence of a finite critical spin does not rely on any specific prescription for the ejection speed $v_{\text{out}}$. Indeed, Eq.~\eqref{eq:inequality} admits a nonempty domain for any fixed $v_{\out}>0$ and a suitable ejection direction. Assuming an Alfv\'enic outflow, as in Ref.~\cite{ComissoAsenjo2021}, one has
\begin{equation}
 v_{\out}\simeq\sqrt{\frac{\sigma_0}{1+\sigma_0}},
 \qquad \sigma_0=\frac{B_0^2}{4\pi\omega_0},
\end{equation}
where $B_0$ is the reconnecting magnetic field strength measured in the upstream plasma rest frame and $\omega_0$ is its proper enthalpy density. Thus, within our prescribed BZ drift model, negative-energy plasmoids are kinematically accessible at arbitrarily small nonzero spins for any fixed $\sigma_0>0$, without requiring increasingly large magnetizations as the spin decreases, in contrast with models based on plunging geodesic motion \cite{Shen:2024sdr}. While the extent of the negative-energy domain depends on $\sigma_0$, the absence of a finite critical spin, which stands as our key result, does not. Figure~\ref{fig:ca-bz-analytic} compares these boundaries in the
$(a,r/M)$ plane with the original Comisso-Asenjo local negative-energy domain. We choose $\xi = \pi/12$ and $\chi =0$.
For the polar ejection plotted in its BZ panel, the leading curves are
\begin{equation}
\begin{aligned}
 \frac{r_*}{M}&=2-\frac38a^2,\\
 \frac{r_{\mathrm{neg}}}{M}
 &=2-\frac{3-2v_{\out}\cos\xi}{8}a^2,
\end{aligned} 
\label{eq:figureedges}
\end{equation}
The BZ panel magnifies $r\simeq2M$ and is restricted to $a\le0.15$.
This figure makes manifest the disappearance of the critical spin threshold for the existence of a region of production of negative-energy states in the present BZ model. This concludes the illustration of the key result presented in this letter. It is now interesting to make several remarks concerning the present model.

\section{Discussion}

The main motivation of this work focused on identifying how the effects of the black hole magnetosphere can modify the spin threshold found earlier in \cite{ComissoAsenjo2021, Chen:2024ggq, Zeng:2026ufr, YuChih:2026ios, Shen:2024sdr, Wan2026} for the generation of negative-energy plasmoids in the slowly rotating Kerr black hole. It reveals that properly accounting for the motion of plasma before reconnection is crucial: turning on the BZ magnetosphere, the upstream plasma is forced to drift in the plane orthogonal to the magnetic field, changing the energetics of the generated plasmoids. As a result, we have shown that the generation of negative-energy plasmoids can occur, within the Kerr ergosphere, at any arbitrary small but non-vanishing spin $a$ and magnetization $\sigma_0$. In particular, this phenomena can be triggered below the previously derived spin threshold $a_{\ast} \sim 0.8$, in agreement with recent GR-PIC numerical simulations \cite{Meringolo2025}. 

At this stage, it is interesting to confront our model to previous analytical models of magnetic reconnection-driven energy extraction \cite{Chen:2024ggq, Zeng:2026ufr, YuChih:2026ios, Shen:2024sdr, Wan2026}. In order to extract energy from the hole, the formation of negative-energy plasmoids is not enough. In general, one needs to impose a second condition, namely that the second positive-energy plasmoid escape to infinity. As shown in \cite{Chen:2024ggq, Zeng:2026ufr, YuChih:2026ios, Shen:2024sdr, Wan2026}, this escape condition tends to raise the value of the associated critical spin threshold. However, we would like here to qualify the need for this second condition. Indeed, in order to extract energy from the black hole, one needs only to impose that the absorbed energy during the process be negative. Crucially, this can be achieved even if the second positive-energy plasmoid gets finally accreted by the hole. To see this, let us call the in-state plasma by $P_{\text{up}}$ (with energy $ \mathcal E_{\text{up}} >0$) and the two oppositely and instantaneously ejected plasmoids by $P_{\pm}$ (with energies $ \mathcal E_{-} < 0$ and $ \mathcal E_{+} >0$). Let us now assume that the plasmoid $P_{+}$ orbits for some time in the magnetosphere and transfers part of its energy to radiation or to the surrounding plasma before getting accreted and ultimately crossing the horizon. Let us denote the amount of energy transferred to radiation or to the surrounding plasma by $\mathcal E^{\text{transf}}_{+}$ and the amount of energy carried by $P_{+}$ when accreted by $\mathcal E^{\text{acc}}_{+}$. Here, all energies refer to total Killing energies rather than specific energies. Assuming that $P_{-}$ retains its initial energy and that no other energy flux crosses the horizon, energy conservation implies $\mathcal E^{\text{acc}}_{+}=\mathcal E_{+}-\mathcal E^{\text{transf}}_{+}$. Then, the net energy balance at the horizon after having accreted both plasmoids is given by
\begin{align}
\Delta \mathcal E |_{r_{+}} = \mathcal E_{-} +\mathcal E^{\text{acc}}_{+} = \mathcal E_{-}+\mathcal E_{+}-\mathcal E^{\text{transf}}_{+}.
\end{align}
Provided that $\mathcal E^{\text{acc}}_{+}<-\mathcal E_{-}$, one still has $\Delta \mathcal E |_{r_{+}} <0$. Equivalently, the energy transferred by $P_{+}$ must satisfy
\begin{align}
\mathcal E^{\text{transf}}_{+}>\mathcal E_{+}+\mathcal E_{-}.
\end{align}
If the total Killing energy is conserved during the reconnection event, with no additional energy channels, the right-hand side is simply $\mathcal E_{\text{up}}$. Assuming that $P_{+}$ still carries non-negative energy when accreted, the required transfer efficiency thus satisfies $1-|\mathcal E_{-}|/\mathcal E_{+}<\mathcal E^{\text{transf}}_{+}/\mathcal E_{+}\leq 1$. Defining the transfer efficiency as $\eta_{\text{transf}}:=\mathcal E^{\text{transf}}_{+}/\mathcal E_{+}$, and assuming that both plasmoids have the same effective rest mass so that the ratio of total energies coincides with that of specific energies, Eq.~\eqref{eq:eout} yields the critical efficiency
\begin{equation}
\begin{aligned}
\eta_{\text{crit}}&=1-\frac{|\mathcal E_-|}{\mathcal E_+}=\frac{2(3-8\kk)}{3-8\kk+v_{\out}|\mathcal A|}+O(a^2),
\end{aligned}
\label{eq:transfer_efficiency}
\end{equation}
with $\eta_{\text{crit}}<\eta_{\text{transf}}\leq 1$. At leading order, this threshold is independent of the black hole spin and tends to zero as $\kk\to3/8$ for fixed nonvanishing $v_{\out}|\mathcal A|$. Crucially, this condition assumes that the transferred energy is not subsequently accreted. Otherwise, only the fraction that remains outside the black hole or escapes to infinity contributes to net extraction. Therefore, under this condition, the process extracts energy from the black hole even if both plasmoids $P_{\pm}$ are ultimately accreted. In particular, part of this energy transfer may occur through radiation emitted by the positive-energy plasmoid before its accretion, potentially giving rise to observable flares powered in part by the rotational energy of the black hole. Therefore, it follows from this discussion that imposing an additional escape condition for the $P_{+}$, as done in most models \cite{Chen:2024ggq, Zeng:2026ufr, YuChih:2026ios, Shen:2024sdr, Wan2026}, is not mandatory. Properly accounting for the transfer of energy from $P_{+}$ to the surrounding plasma is beyond the scope of the present work. Nevertheless, it opens another window for the construction of energy extraction via magnetic reconnection.  

Finally, it would be interesting to further explore how to go beyond the limits of the present model. In particular, it would interesting to search if a simple exact model for the contribution $F^{\text{MR}}_{\mu\nu}$ responsible for the magnetic reconnection can be found on Kerr, similarly to the exact solutions known in flat spacetime such as the Craig-Henton model \cite{CraigHenton1995}.  If so, one could in principle lower the numbers of free parameters in the present model. We leave this for future works.

\begin{acknowledgments}
J.BA would like to thank the members of the group of the San Sebastian University in Santiago for their kind hospitality during his visit where this project was worked out as well as Felipe Asenjo for discussions. We acknowledge the use of AI services to prepare figures, performing independent cross-checks with our own results and for type-setting. The authors assume full responsibility for the accuracy and integrity of the scientific content presented in this article.
\end{acknowledgments}

\clearpage
\onecolumngrid
\appendix

\begin{center}
{\large\bfseries Supplementary Material}
\end{center}

\section{BZ field and Znajek regularity}\label{app:bz}
In this appendix, we gather the details on the Blandford-Znajek solution used in this work.  For the northern hemisphere of the stationary, axisymmetric, degenerate split monopole, the
magnetic flux function is given by
\begin{equation}
\begin{aligned}
\Psi&=A_\phi=\Psi_0(1-\cos\theta)+O(a^2),\\
F&=d\Psi\wedge(\dd\phi-\Omega_F\dd t)+F_{r\theta}\,\dd r\wedge\dd\theta,
\end{aligned}
\label{eq:bzF}
\end{equation}
where $\Omega_F=a/(8M)+O(a^3/M)$.  The first term gives
$F_{\theta\phi}=\Psi_{,\theta}$ and
$F_{\theta t}=-\Omega_F\Psi_{,\theta}$ while the second encodes the toroidal
field and is fixed at the horizon by the current/regularity condition.
With $\nabla_\nu F^{\mu\nu}=4\pi J^\mu$ and
$\mathcal I\equiv\sqrt{-g}\,F^{\theta r}$, one has
\begin{equation}
\begin{aligned}
4\pi\sqrt{-g}\,J^r&=-\partial_\theta\mathcal I,\\
4\pi\sqrt{-g}\,J^\theta&=\partial_r\mathcal I.
\end{aligned}
\label{eq:bzcurrent}
\end{equation}
In order to derive horizon regularity without using a singular horizon ZAMO, onne can 
introduce ingoing Kerr coordinates which read
\begin{equation}
\begin{aligned}
\dd v&=\dd t+\frac{r^2+a_K^2}{\Delta}\dd r,\\
\dd\tilde\phi&=\dd\phi+\frac{a_K}{\Delta}\dd r.
\end{aligned}
\label{eq:ingoing}
\end{equation}
where $a_K=aM$.
Substituting these in \eqref{eq:bzF} yields
\begin{equation}
\begin{split}
 F={}&\dd\Psi\wedge(\dd\tilde\phi-\Omega_F\dd v)+\left[F_{r\theta}^{\mathrm{BL}}
 -\frac{\Omega_F(r^2+a_K^2)-a_K}{\Delta}\Psi_{,\theta}\right]
 \dd r\wedge\dd\theta.
\end{split}
\label{eq:ingoingF}
\end{equation}
The key point is that the square bracket must remain finite on the future horizon.  With
$\Omega_H=a_K/(r_+^2+a_K^2)$, this implies
\begin{equation}
\begin{aligned}
 F_{r\theta}^{\mathrm{BL}}
 &=-\frac{(r_+^2+a_K^2)(\Omega_H-\Omega_F)}{\Delta} \Psi_{,\theta}+O(\Delta^0),\\
 \mathcal I_H&=(\Omega_H-\Omega_F) \frac{(r_+^2+a_K^2)\sin\theta}{\Sigma_+}\Psi_{,\theta}.
\end{aligned}
\label{eq:znajek}
\end{equation}
Thus the Boyer--Lindquist divergence of $F_{r\theta}$ is compatible
with a smooth electromagnetic two-form on the future horizon.  For the
slowly rotating split monopole, one has
$\mathcal I_H=(a\Psi_0/8M)\sin^2\theta+O(a^3)$.

We can now compute the electric ad magnetic decomposition w.r.t a ZAMO. The orthonormal fields measured by the ZAMO are, to this order,
\begin{equation}
\begin{aligned}
 B^{\hat r}&=\frac{\Psi_{,\theta}}{\sqrt\Sigma\,\varpi},\\
 E^{\hat\theta}&=\frac{\omega-\Omega_F}{\alpha\sqrt\Sigma}\Psi_{,\theta},\\
 B^{\hat\phi}&=\frac{\sqrt\Delta}{\Sigma}F_{r\theta}.
\end{aligned}
\label{eq:EBfromF}
\end{equation}
Since $r_X-r_+=O(a^2M)$, the singular term in
\eqref{eq:znajek} dominates $F_{r\theta}$ throughout the stretched
layer, and \eqref{eq:EBfromF} gives
\begin{align}
B^{\hat\phi} & =-E^{\hat\theta}+O(a^2B^{\hat r}) \\
E^{\hat\theta} & =-\varpi(\Omega_F-\omega)B^{\hat r}/\alpha
\end{align}
which yields 
\begin{align}
\frac{B^{\hat\phi}}{B^{\hat r}}=v_F=-\frac{1}{2\sqrt{1-2\kk}}+O(a^2)
\end{align}
which is Eq.~\eqref{eq:fields} in the main text.  The future-horizon boundary condition
therefore fixes the plunging sign of the minimal BZ drift:
\begin{align}
v_D^{\hat r}=- \frac{v_F^2}{(1+v_F^2)} <0
\end{align}

\end{document}